\documentstyle[12pt,fullpage,psfig]{article}

\begin{document}

\title{Comparison of 2D and 3D models of flow structure in
semidetached binaries}

\author{D.V.Bisikalo$^1$, A.A.Boyarchuk$^1$,\\
        V.M.Chechetkin$^2$, O.A.Kuznetsov$^2$,
        D.Molteni$^3$\\[5mm]
$^1$ Institute of Astronomy of Russian\\
Academy of Sciences, Moscow\\
$^2$ Keldysh Institute of Applied Mathematics, Moscow\\
$^3$ Dipartimento di Scienze Fisiche ed Astronomiche,\\
Universit\`a di Palermo}

\date{}

\maketitle

\begin{abstract}

{\bf Abstract}---We present the results of systematic comparison
of 2D and 3D numerical models of mass transfer in semidetached
binaries. It is shown that only for case of $\gamma \sim 1$
(near-isothermal case) the obtained 2D and 3D solutions are
qualitatively similar.  For higher value of $\gamma$ the 3D flow
structure is drastically changed and for $\gamma=1.2$ the
accretion disk is not formed in the system. Numerical results
show that for case of $\gamma=1.2$ the 2D and 3D solutions are
different.

We discuss the spiral-shaped shock waves obtained in numerical
models. It is shown that these shocks are not intrinsic spirals
and are caused by collisions of the gaseous flows in the system.

\end{abstract}

\section*{INTRODUCTION}

The semidetached binaries are the interacting stars, where one
component fills its Roche lobe and mass transfer between
components of the system occurs through the vicinity of inner
Lagrangian point $L_1$. The semidetached binaries  --
cataclysmic variables, low-mass X-ray binaries, supersoft X-ray
sources, etc. --  show a lot of bright evidences of a complex
flow structure that should be interpreted. Since the pioneer
work by Prendergast [\ref{PRE60}] the gasdynamical
treatment of the flow patterns in semidetached binaries have
been made by many workers. The numerical simulations were
conducted both in 2D (see, e.g.,
%[\ref{SAW86}, \ref{SAW87},
%\ref{SPRU87}, \ref{ROZ89}, \ref{TAAM91}, \ref{BLON95},
%\ref{MUR96}]
[\ref{SAW86}--\ref{MUR96}]) and in 3D
%([\ref{NAG91}, \ref{HIR91},
%\ref{MOL91}, \ref{SAW92}, \ref{LANZ92}, \ref{BELV93},
%\ref{MEG93}, \ref{LANZ94}, \ref{BIS97A}, \ref{BIS97B},
%\ref{YUK97}, \ref{BIS98}])
([\ref{NAG91}--\ref{BIS98}])
models.  For a long time the 3D
numerical simulation was limited by computer power therefore the
2D models were the preferred ones for the analysis of the flow
structure. A lot of interesting results were obtained using the
two-dimensional approach and a set of bright ideas were based on
this numerical simulations.  The known restrictions of the 2D
approach cause the question of the validity of used models and
of reliability of obtained results.  The question of the 2D
approach applicability became especially important since 1994
when Fridman \& Khoruzhii [\ref{FRID94}] have analytically shown
that in the accretion disk the 2D approach is valid only for two
specific cases (for isothermal disk under the action of external
gravitational field, and for self-gravitating disk with heats
ratio $\gamma=2$), while for all other cases the only 3D models
are valid.

It is evident that the direct way to estimate the applicability
of 2D models is the comparison of 2D and 3D results.
Unfortunately, the comparison between results obtained earlier
in 2D and 3D models by previous authors is rather difficult
because they have used different numerical technics and have
considered  different binary parameters. To be confirmed in the
results of a comparison it is necessary to perform a set of 2D
and 3D simulation using the same numerical technique for the
same binary system and for the same boundary conditions. The
only one attempt to make such comparison is due to Sawada and
Matsuda [\ref{SAW92}], but their 3D calculations were rather
crude due to bad spatial resolution and small time of
calculations (they have stopped the run at the time less than a
half of orbital period, i.e. when the steady-state regime was
not established). Therefore, while the authors have declared
the qualitative similarity of 2D and 3D results this conclusion
was not well argued.

In the present work, for the first time, we have made the
systematic comparison of the 2D and 3D models. We perform two-
and three-dimensional hydrodynamic simulation of the flow of
inviscid adiabatic gas in a non-magnetic semidetached binary.
The description of the used models are presented in Section 2.
The 3D runs is very time-consuming, nevertheless in 3D model we
use the rather fine grid (only 2 times coarser than in 2D model)
and conduct the runs up to the steady-state flow regime (up to
10 orbital periods) which allows us to make a reliable
comparison with 2D results. In Section 3 it is shown that the
qualitative similarity of 2D and 3D results exists only for the
case of the $\gamma \sim 1$, while for higher $\gamma$ the
solutions are different. In Section 4 we discuss the changes of
the flow structure caused by different values of $\gamma$.

One of the most interesting numerical results obtained in
previous works is the presence of two-armed spiral shocks in
accretion disk. In a set of numerical and analytical works (see,
e.g.,
%[\ref{SAW86}, \ref{SAW87}, \ref{SPRU87}, \ref{ROZ89},
%\ref{SAW92}])
[\ref{SAW86}--\ref{ROZ89},\ref{SAW92}])
these shocks have been considered as intrinsic
spiral shocks caused by the tidal disturbances of accretion
disk. The set of 2D and 3D runs, conducted in the present work,
shows the existence of spiral-shape shocks in the semidetached
binaries, but as it follows from the analysis of results these
shocks are caused by collision of gas fluxes in the system. In
Section 5 we discuss the proves of the "flow crossing" nature of
obtained spiral-shaped shocks. Our conclusions follow in Section
6.

\section{THE MODEL DESCRIPTION}

To compare results of this work with previous ones we study the
non-magnetic binary system with the same parameters as in
[\ref{SAW86}]. We consider a semi-detached binary with
components mass ratio of unity. The mass-losing star fills up
its critical Roche lobe that causes the mass transfer between
components of the system.

To describe the gas flow in the binary we use the standard Euler
equations in the reference frame corotating with the binary
system. The 3D simulations are made in Cartesian coordinates
($x,y,z$) and the equations have the form:

$$
\frac{\partial \rho}{\partial t}
+\frac{\partial \rho u}{\partial x}
+\frac{\partial \rho v}{\partial y}
+\frac{\partial \rho w}{\partial z}=0
$$

$$
\frac{\partial\rho u}{\partial t}
+\frac{\partial}{\partial x}(\rho u^2+P)
+\frac{\partial\rho u v}{\partial y}
+\frac{\partial\rho u w}{\partial z}
= -\rho\frac{\partial\Phi}{\partial x}+2\Omega v\rho
$$

$$
\frac{\partial\rho v}{\partial t}
+\frac{\partial\rho u v}{\partial x}
+\frac{\partial}{\partial y}(\rho v^2+P)
+\frac{\partial\rho v w}{\partial z}
= -\rho\frac{\partial\Phi}{\partial y}-2\Omega u\rho
$$

$$
\frac{\partial\rho w}{\partial t}
+\frac{\partial\rho u w}{\partial x}
+\frac{\partial\rho v w}{\partial y}
+\frac{\partial}{\partial z}(\rho w^2+P)
= -\rho\frac{\partial\Phi}{\partial z}
$$

$$
\frac{\partial\rho E}{\partial t}
+\frac{\partial\rho u h}{\partial x}
+\frac{\partial\rho v h}{\partial y}
+\frac{\partial\rho w h}{\partial z}
= -\rho u\frac{\partial\Phi}{\partial x}
-\rho v\frac{\partial\Phi}{\partial y}
-\rho w\frac{\partial\Phi}{\partial z}\,.
$$

\noindent Here $\rho$ -- density; ${\bf v}(x,y,z)=(u,v,w)$
-- velocity vector; $P$  -- pressure; $E$  -- specific full
energy $E=\varepsilon+{\bf v}^2/2$; $\varepsilon$ -- internal
specific energy; $h$ -- specific full enthalpy
$h=\varepsilon+P/\rho+{\bf v}^2/2$; $\Phi$ -- Roche potential

$$
\Phi({\bf r}) = -\frac{G
M_1}{|{\bf r}-{\bf r}_1|}-\frac{G M_2}{|{\bf r}-{\bf
r}_2|} -\frac{1}{2}\Omega^2\left({\bf r}-{\bf
r}_c\right)^2\,,
$$
where ${\bf{r}}_1$, ${\bf{r}}_2$ -- the centers of
components of the system; and ${\bf{r}}_{\rm c}$ -- the
center of mass of the system; $M_1$, $M_2$ -- masses of
components; $\Omega=2\pi/P_{\rm orb}$ -- angular velocity of the
system.

The 2D simulations are made in cylindrical coordinates
($r,\phi$) and the equations have the form:

$$
\frac{\partial \rho}{\partial t}
+ \frac{1}{r}\frac{\partial r\rho u}{\partial r}
+ \frac{1}{r}\frac{\partial\rho v}{\partial \varphi} = 0
$$

$$
\frac{\partial\rho u}{\partial t}
+\frac{1}{r}\frac{\partial}{\partial r}(r\rho u^2+rP)
+\frac{1}{r}\frac{\partial\rho u v}{\partial \varphi}
=\frac{P}{r}
-\rho\frac{\partial\Phi}{\partial r}
+\rho\frac{v^2}{r}
+2\Omega v\rho
$$

$$
\frac{\partial\rho v}{\partial t}
+\frac{1}{r}\frac{\partial r\rho u v}{\partial r}
+\frac{1}{r}\frac{\partial}{\partial\varphi}(\rho v^2+P)
=-\frac{\rho}{r}\frac{\partial\Phi}{\partial\varphi}
-\rho\frac{u v}{r}
-2\Omega u\rho
$$

$$
\frac{\partial\rho E}{\partial t}
+\frac{1}{r}\frac{\partial r\rho u h}{\partial r}
+\frac{1}{r}\frac{\partial\rho v h}{\partial \varphi}
=-\rho u \frac{\partial\Phi}{\partial r}
-\rho\frac{v}{r}\frac{\partial\Phi}{\partial\varphi}\,,
$$
where ${\bf v}(r,\varphi)=(u,v)$ -- velocity vector.

Both for 2D and 3D cases we used the ideal gas equation of state
$P=(\gamma-1)\rho\varepsilon$ where $\gamma$ is the ratio of
specific heats. Following to procedure suggested by
[\ref{SAW86}] the 2D and 3D equations are adimensionalized
using the separation $A$ of the two stars as the length scale,
the reciprocal of the orbital angular velocity ($\Omega^{-1}$)
as the time-scale and the density on the surface of mass-losing
star as the density scale. The used velocity scale is $A\Omega$.

To specify the physical boundary conditions on the Roche lobe of
the mass-losing star we have to settle the density ($\rho_0$),
sound velocity ($c_0$), and gas velocity vector ${\bf v}_0$. It
should be noted however that: i) the boundary value of density
on the surface of mass-losing star has no influence on the
solution, due to scaling of the system of equations with respect
to $\rho$ (with simultaneous scaling of $P$), and ii) in
adimensioned approach the used value of sound velocity can
correspond to arbitrary value of the temperature in dependence
of adopted physical values of $M_1$, $M_2$ and $A$. Therefore to
compare results with previous ones we have adopted the same
adimensioned gas parameters at the surface of mass-losing star
as in the work by [\ref{SAW86}]:

   - the value of density $\rho_0$ is equal to 1;

   - sound speed $c_0$ is equal to 0.15;

   - the gas is assumed to be ejected perpendicularly to the Roche
surface with velocity $u_0 = 0.0125$.

The numerical boundary conditions are built using the standard
procedure of solving of Riemann problem between two regions --
one on the surface of the star and other corresponding to the
nearest computational cell (see, e.g., [\ref{SAW86},
\ref{SAW92}]).

The radius of the compact star is assumed to be equal to
$0.01A$.  Free outflow conditions are imposed on the accretor
and on the outer boundary.

To solve the equations we use the TVD Roe scheme [\ref{ROE86}]
of first order of approximation with monotonic flux limiters in
form of Osher [\ref{CHAK85}] that makes the scheme of third
order of approximation. We also include Einfeldt modification of
Roe scheme [\ref{EIN88}] to make the scheme more stable.

The computation domain for 2D case is a circle $r\leq 3A$
(except the Roche lobe filled by mass-losing star and a circle
representing the accretor) covered by the non-uniform grid
containing $231 \times 240$ gridpoints in $r$ and $\phi$
directions respectively. For 3D case the computation domain is a
parallelepipedon $[-A..2A]\times[-A..A]\times[0..A]$ (with the
same exceptions) covered by the non-uniform grid containing
$132\times 107\times 30$ gridpoints in $X$, $Y$ and $Z$
directions respectively. In both 2D and 3D the grids are
condensed in the vicinity of accretor. In 3D runs symmetry about
the equatorial plane is assumed.  The numerical simulations are
conducted up to nearly 10 orbital periods both for 2D and 3D
cases.

\begin{figure}[t]
\centerline{\hbox{\psfig{figure=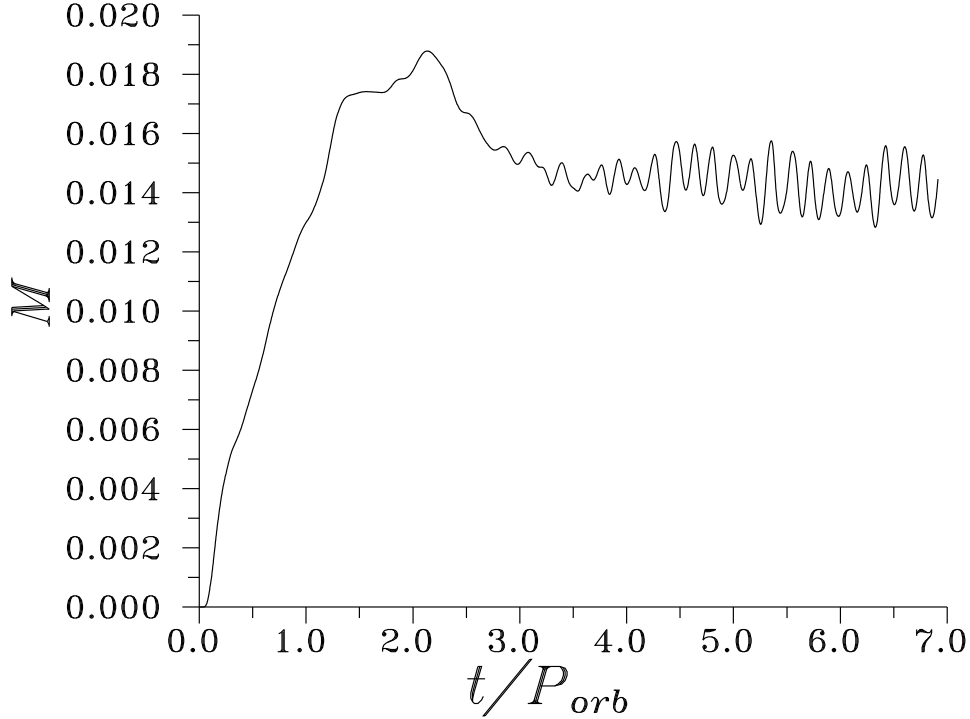,width=5in}}}
\caption{
Surface mass ($\int \rho dS$) contained
in the circle with radius 0.3$A$ around accretor for 2D model
with $\gamma=1.2$ as a function of time.}
\end{figure}

\begin{figure}[t]
\centerline{\hbox{\psfig{figure=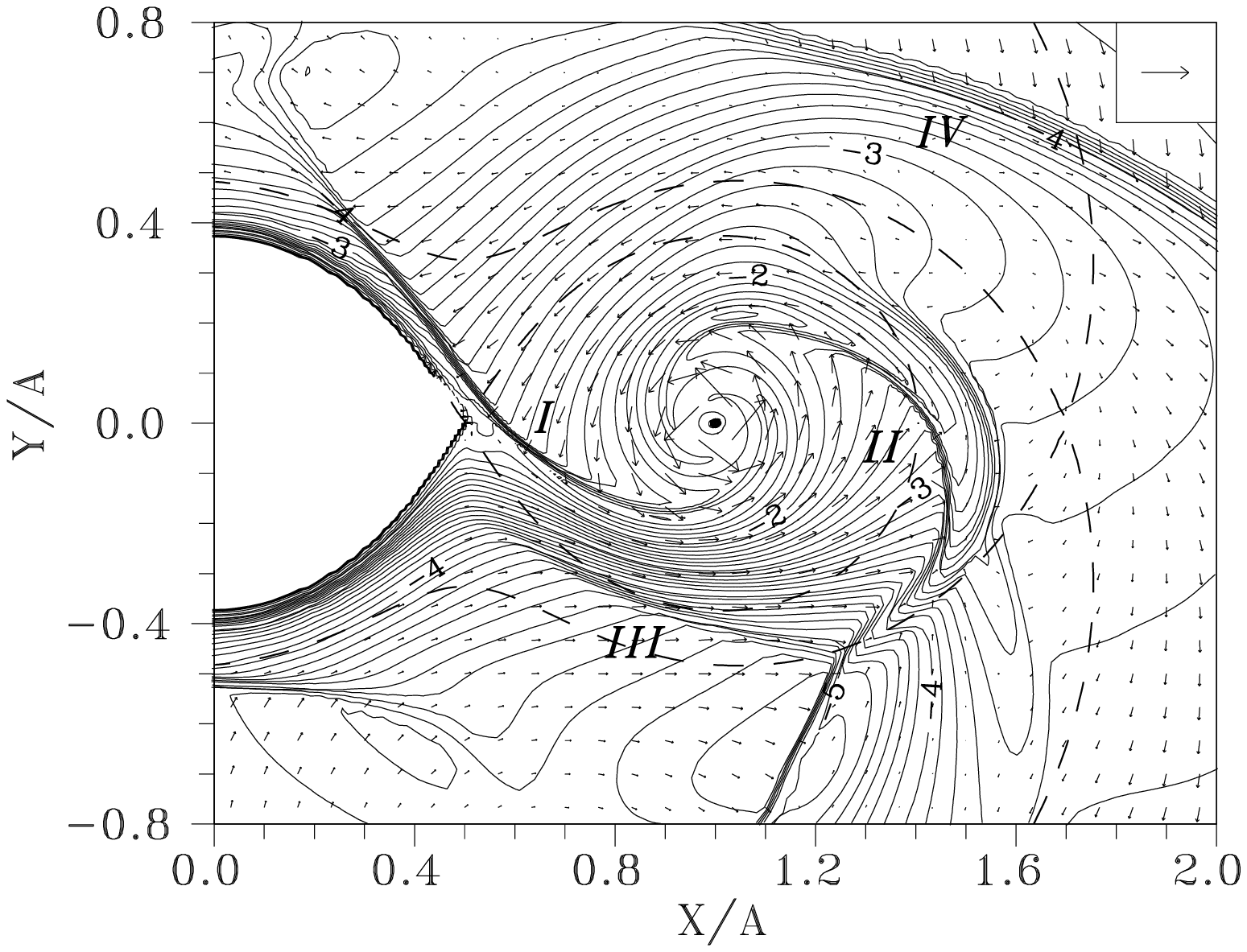,width=6in}}}
\caption{
Pressure isolines and velocity vectors in
the equatorial plane of the system for the 2D run (case
$\gamma=1.01$) between $10^{-5}..0.2$ with constant increment of
$\lg P$.  Digits on isolines denote values of $\lg P$.  Roche
equipotentials are shown by dashed lines.  Shocks are marked by
roman digits $I$, $II$, $III$, $IV$. The accretor is marked by
the filled circle.  Vector in the upper right corner corresponds
to the value of velocity of $3A\Omega$.}
\end{figure}

\begin{figure}[t]
\centerline{\hbox{\psfig{figure=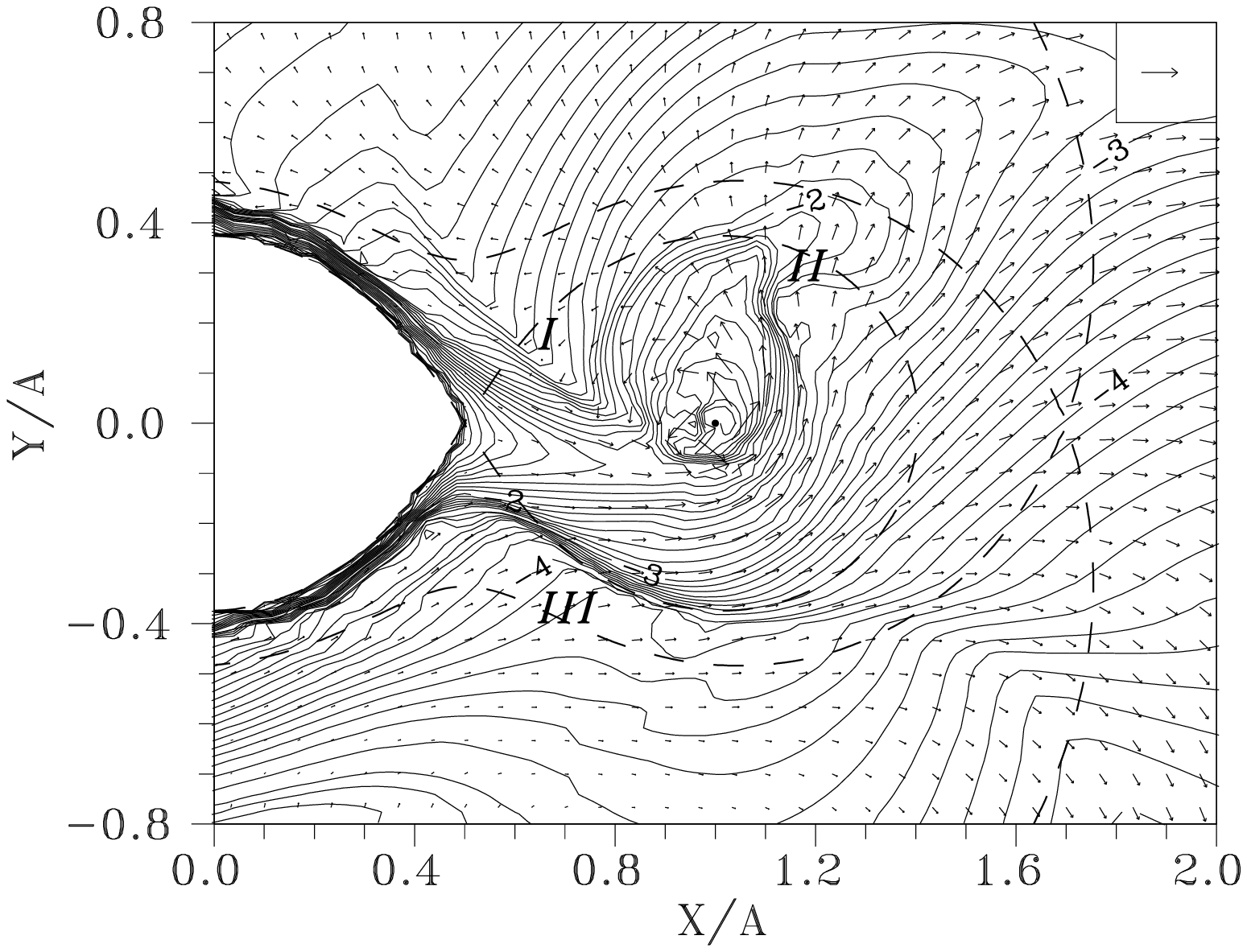,width=6in}}}
\caption{
Pressure isolines and velocity vectors in
the equatorial plane of the system for the 3D run (case
$\gamma=1.01$) between $10^{-5}..0.2$ with constant increment of $\lg
P$. Digits on isolines denote values of $\lg P$.  Roche
equipotentials are shown by dashed lines. Shocks are marked by
roman digits $I$, $II$, $III$.  The accretor is marked by the
filled circle. Vector in the upper right corner corresponds to
the value of velocity of $3A\Omega$.}
\end{figure}

\begin{figure}[t]
\centerline{\hbox{\psfig{figure=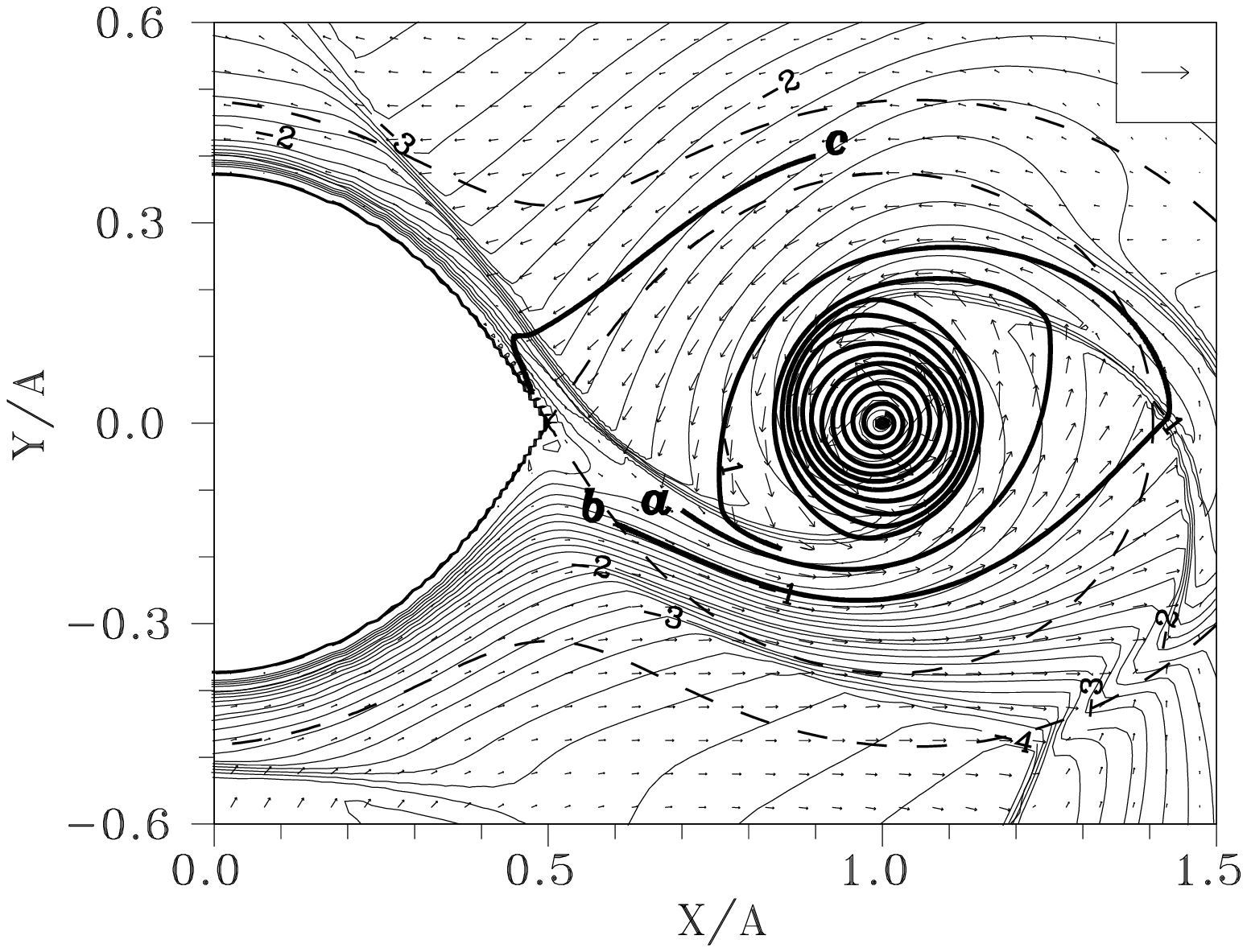,width=6in}}}
\caption{
Density isolines and velocity vectors in
the equatorial plane of the system for the 2D run (case
$\gamma=1.01$) between $10^{-5}..5$ with constant increment
of $\lg \rho$. Digits on isolines denote values of $\lg \rho$.
Different flowlines are marked by letters "$a$", "$b$", "$c$".
Vector in the upper right corner corresponds to the value of
velocity of $3A\Omega$.}
\end{figure}

\begin{figure}[t]
\centerline{\hbox{\psfig{figure=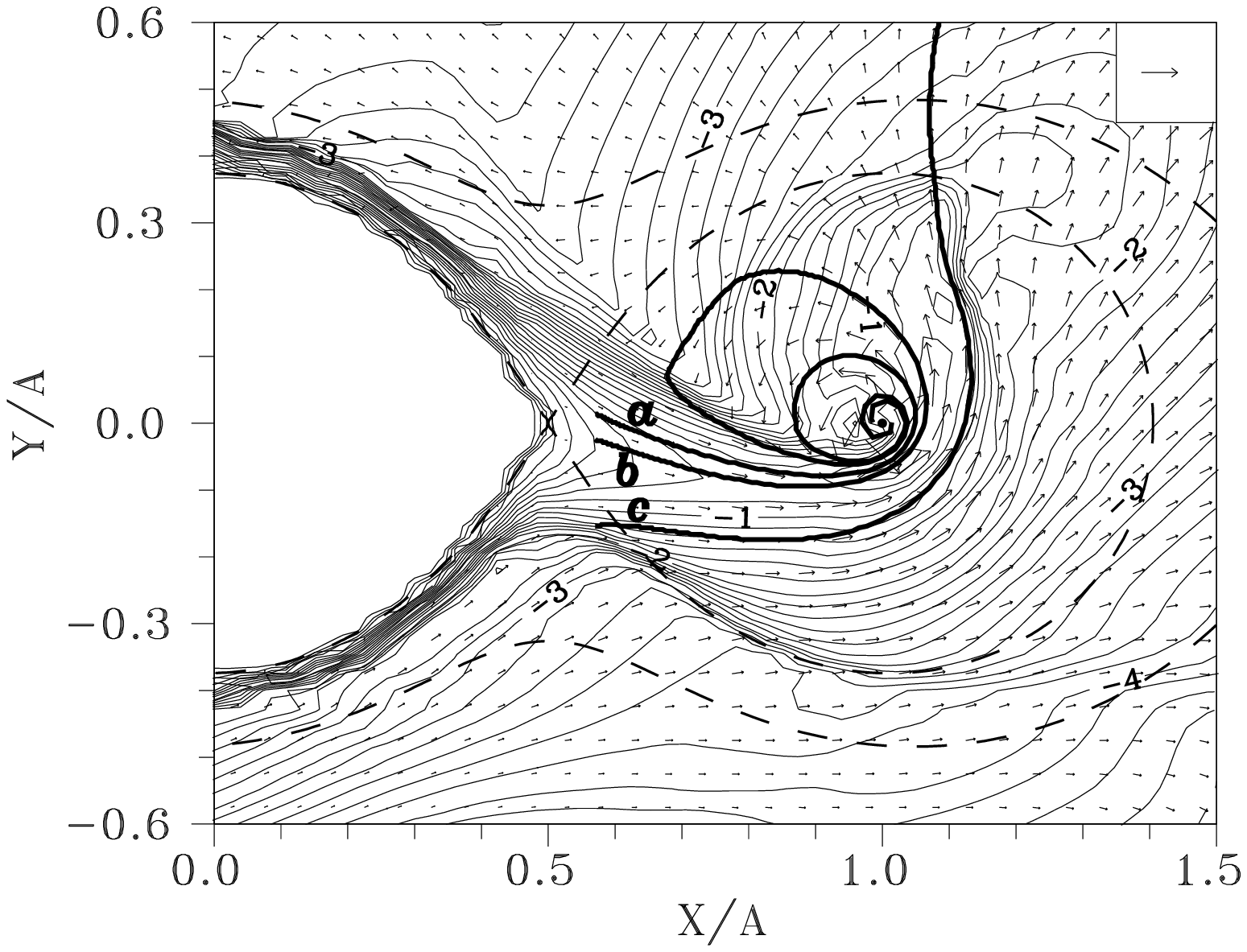,width=6in}}}
\caption{
Density isolines and velocity vectors in
the equatorial plane of the system for the 3D run (case
$\gamma=1.01$) between $10^{-5}..1$ with constant increment
of $\lg \rho$. Digits on isolines denote values of $\lg \rho$.
Different flowlines are marked by letters "$a$", "$b$", "$c$".
Vector in the upper right corner corresponds to the value of
velocity of $3A\Omega$.}
\end{figure}

\section{COMPARISON OF 2D AND 3D RUNS}

To perform the systematic comparison of the 2D and 3D models we
conduct gasdynamical simulations for the same semidetached
binary.  The boundary conditions (density, gas velocity, and
sound speed) at the surface of mass-losing star are kept
constant in different runs, while the ratio $\gamma$ of specific
heats is a parameter to be varied. Four cases -- two runs for
$\gamma=1.01$ and $\gamma=1.2$ both in 2D and 3D models -- are
conducted. To study the influence of the numerical viscosity on
the solution we have also conducted 2D runs (both for
$\gamma=1.01$ and $\gamma=1.2$ cases) with the same coarse grid
as we used for 3D model. The coarse-grid and fine-grid 2D
solutions show the same structure of the gaseous flows in the
system except the additional viscous broadening of shocks,
therefore below we will describe the fine-grid 2D results.

Before describing the results of comparison it should be noted
that the 3D solutions are a steady-state ones. In all 3D runs
there are no changes of the flow structure for the time
exceeding $\sim 1.5$ orbital periods. The analysis of average
characteristics of the obtained solutions shows that all
parameters of the flow are stable in all 3D runs. For 2D case
the solutions do not reach the steady-state and even for large
evolution time quasi periodic changes of the flow structure are
observed. The confirmation of the non steady-state behavior of
2D solution can be got from the analysis of the  average
characteristics of the flow, e.g. from the consideration of the
time-dependence of the mass variation in the fixed volume around
accretor. In Figure~1 the surface mass ($\int \rho dS$)
contained  in the circle with radius 0.3$A$ around accretor for
$\gamma=1.2$ run is presented.  It is seen that for the 2D
solution the mass changes quasi periodically for all time. The
observed oscillations are probably caused by restrictions of the
2D approach.  The inconsistency of the mass accretion rate with
the rate of mass transfer from the mass-losing star to accretion
disk lead to non-steady-state behavior of the disk. Due to the
absence of the $Z$ motion of the gas in the 2D models the
accumulated matter of the disk can exceed the limitation imposed
by the gasdynamical structure of the flow. Then the excess of
the matter tries to leave the system and as a consequence
changes the gaseous flow structure.  To compare the 2D and 3D
models we extract the most typical characteristic features of
the 2D flow patterns.

\subsection{The case of $\gamma$=1.01}

The general structure of gaseous flows in the equatorial plane
of considered binary system for $\gamma=1.01$ case is presented
in Figures~2--5. In Figures~2 and 3 the isolines of pressure and
velocity vectors in the region $[0..2A] \times [-0.8A..0.8A]$
are shown for 2D and 3D cases, respectively. The isolines of
density, velocity vectors, and flowlines for the same runs in
the region $[0..1.5A] \times [-0.6A..0.6A]$ are presented in
Figures~4 and 5. Figures are presented in rotating systems of
coordinates with counter-clockwise directions of rotation and
angular velocity $\Omega$. For 3D case arrows show the 3D
velocities projected onto the slice plane.

The analysis of the results presented in Figures~2 and 3 shows
that for considered near-isothermal case the flow structure
obtained in 2D and 3D models have a set of common features: i)
the accretion disk is formed; ii) there is a circumbinary
envelope in the system; iii) the spiral-shaped shocks labeled by
markers $I$ and $II$ are formed; iv) the interaction of the gas
of the circumbinary envelope with the stream causes the
formation of the shock wave labeled by marker $III$.

To make a more detailed analysis of the structures of gaseous
flows let us consider the density fields and velocity vectors in
the equatorial plane presented in Figures~4 and 5. In each
figure three flowlines, labeled by markers "$a$", "$b$" and
"$c$", are shown as well. These flowlines illustrate directions
of matter flow in the system. The analysis of results presented
in Figures~4 and 5 shows that for near-isothermal case the flow
structure has the same characteristic features as it was
discussed in our previous works [\ref{BIS97A}, \ref{BIS97B},
\ref{BIS98}].  The matter of the stream is redistributed into
two parts: i) the first part (flowline "$a$") forms a
quasi-elliptic accretion disk; ii) the second part (gas flows
labeled by markers "$b$" and "$c$") forms the circumbinary
envelope of the system. As it is seen from the figures the part
of the circumbinary envelope makes a full revolution around the
accretor (flowline "$b$") and mixes with the matter of the
stream.  This matter does not belong to the disk, because after
mixing with the stream this gas can be considered as the stream
itself. Part of the stream which does not interact with the disk
immediately (flowlines "$c$" in Figs~4, 5) forms the outer part
of the circumbinary envelope.

The presented flow structure shows that in our self-consistent
solution the accretion disk is the part of the stream matter
gravitationally captured by accretor. It is evident that the gas
flow of the stream (deflected under the action of the gas of the
circumbinary envelope) approaches the disk along the tangent
line and does not cause the formation of the traditional "hot
spot". On the other hand, the interaction between the gas of
circumbinary envelope and the stream results in formation of an
intensive shock wave $I$, located along the stream edge turned
towards orbital movement. As it is seen from the figures this
shock arises not only at the stream edge, but also along the
surface of Roche lobe, where the circumbinary envelopes interact
with the mass-losing star. The interaction of the circumbinary
envelope with the stream also leads to formation of rather
intensive shock $III$ at the other side of the stream, and of
weak shock $II$, located at the boundary between the stream
deviated by accretor and the gas of circumbinary envelope.

All characteristic feature for considered near-isothermal state
are present both in 2D and 3D models. Nevertheless, as it
follows from a comparison of the results presented in
Figures~2--5 there are also some quantitative differences in the
flow structures.  In particular, in 3D case the accretion disk
is rather small and the stream of the matter leaving $L_1$ goes
close to the accretor, while in 2D case the disk is large and
more circular and the stream goes far from the accretor. This
fact leads to redistribution  of the gas fluxes in the system
and, as a consequence, to small changes of the positions of the
shocks $I$, $II$, $III$. Moreover, in 2D simulation we can see
that the orbital motion of the accretor in the gas of the outer
part of circumbinary envelope cause the formation in the system
of the typical bow-shock labeled by $IV$, while in 3D case this
bow shock does not arise. It is also seen that in 3D simulation
the shock intensities are smaller than in 2D model (e.g. the
intensity of shock $II$ is so small that we can see it only in
simulations using the scheme of the third order of
approximation, while in the run of first order of approximation
this shock disappear).  The reasons of less intensities of the
3D shocks are rather evident and are due to the possibility of
the gas motion in the $Z$-direction.

From the above description we may conclude that in spite of some
quantitative differences in the results of 2D and 3D simulations
the qualitative characteristics of the flow structures are
similar. In turn, it means that the 2D models give a rather
correct qualitative description of the gas flow structure for
near-isothermal case.

\subsection{The case of $\gamma$=1.2}

In Figures~6--9 the results of 2D and 3D simulations are
presented for the case $\gamma$=1.2. As in Figures~2, 3 the
isolines of pressure and velocity vectors are shown in the same
region of equatorial plane for 2D (Fig.~6) and 3D (Fig.~7)
cases. In Figures~8, 9 the isolines of density, velocity
vectors, and flowlines are also presented.

Analysis of 2D results conducted for $\gamma=1.2$ (Figs~6, 8)
allows to reveal the characteristic features of the flow
structure.  As it is seen from the Figures~6 and 8 the part of
the stream (flowline "$a$" in Fig.~8) forms the accretion disk
in the system. The higher value of adopted $\gamma$ leads to
less effective cooling of the disk (in comparison with
$\gamma=1.01$ case), and, as a consequence, to formation of
rather large disk.  The stream-disk interaction is shock free.
As in the $\gamma=1.01$ case the forming circumbinary envelopes
(flowlines "$b$", "$c$" in Fig.~8) interacts with stream and the
surface of mass-losing star and leads to formation of the shocks
$I$--$III$. The bow shock $IV$ caused by orbital motion of the
accretor in the circumbinary envelope is also present in the
calculated flow structure. The most important changes are
related with the positions of spiral-shaped shocks $I$, and
$II$. For considered case of $\gamma=1.2$ the shocks $I$ and
$II$ go through the disk up to accretor.  Assuming that these
shocks are caused by the interaction of the circumbinary
envelope with the stream (as in the case of $\gamma=1.01$) their
behavior is somehow predictable. For shock formed due to
colliding of flow with obstacle the distance between obstacle
and shock increases with increasing of $\gamma$
[\ref{LANDAW}]. So the shock $I$ has to
form at some distance from the stream, while for $\gamma=1.01$
it is placed on the edge of the stream.  If the shock is in the
disk it has to propagate to the inner part due to circular
motion of the gas around accretor. In considered case the shocks
are strong enough and stopped directly at the accretor surface.
The same explanation can be applied for describing the changing
of the shock $II$ position for $\gamma=1.2$ case.

\begin{figure}[t]
\centerline{\hbox{\psfig{figure=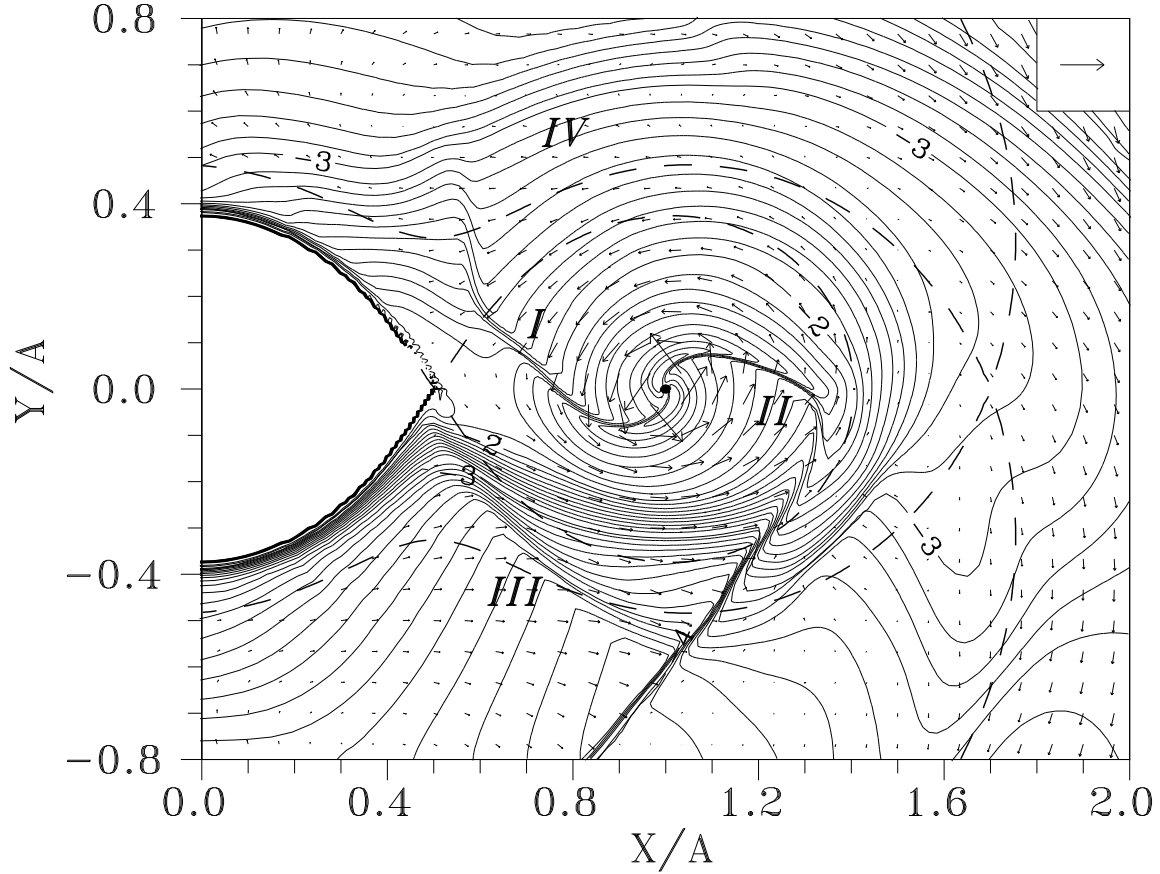,width=6in}}}
\caption{
Pressure isolines and velocity vectors in
the equatorial plane of the system for the 2D run (case
$\gamma=1.2$) between $10^{-5}..0.2$ with constant increment of
$\lg P$.  Digits on isolines denote values of $\lg P$.  Roche
equipotentials are shown by dashed lines.  Shocks are marked by
roman digits $I$, $II$, $III$, $IV$. The accretor is marked by
the filled circle.  Vector in the upper right corner corresponds
to the value of velocity of $3A\Omega$.}
\end{figure}

\begin{figure}[t]
\centerline{\hbox{\psfig{figure=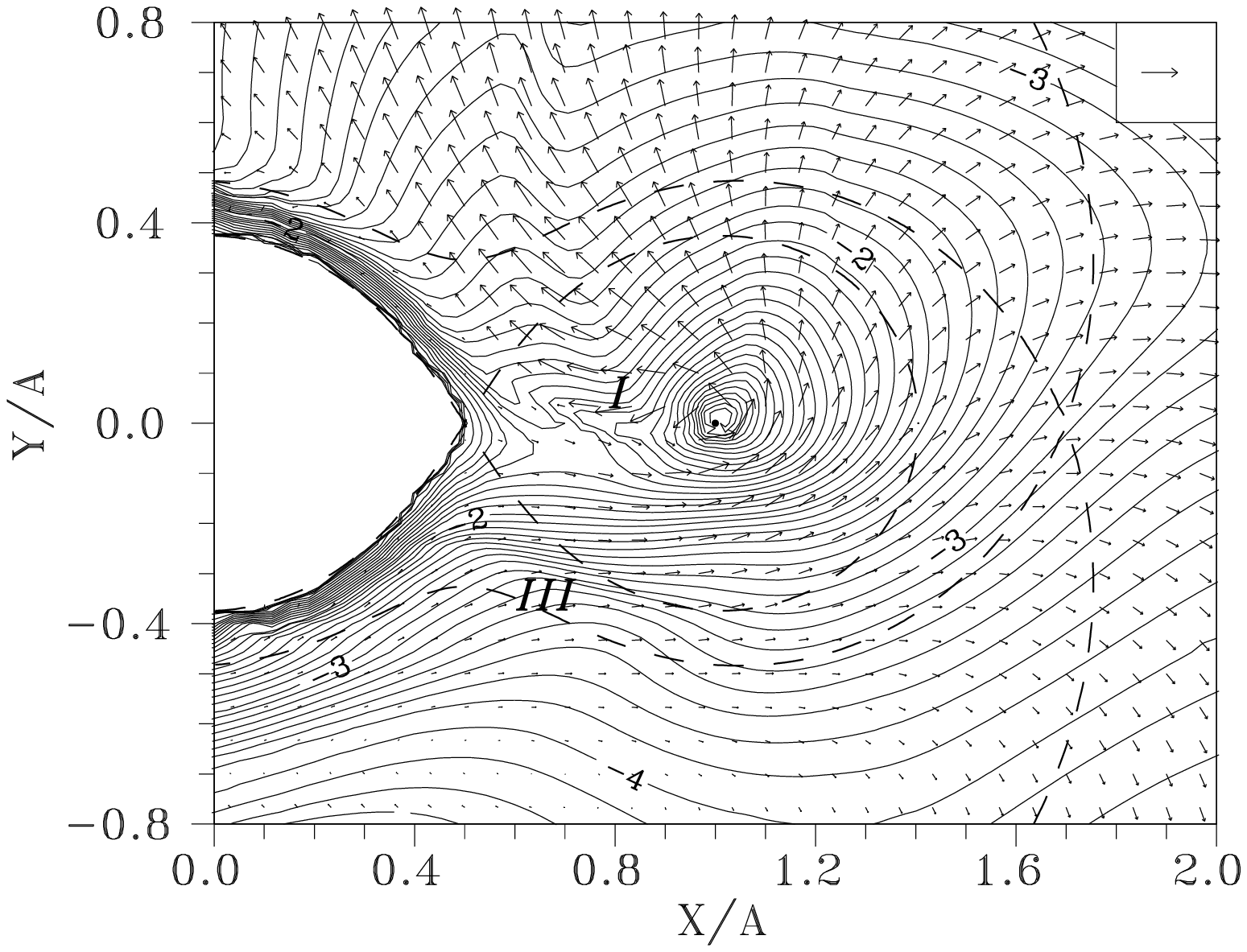,width=6in}}}
\caption{
Pressure isolines and velocity vectors in
the equatorial plane of the system for the 3D run (case
$\gamma=1.2$) between $10^{-5}..0.2$ with constant increment of $\lg
P$.   Digits on isolines denote values of $\lg P$.  Roche
equipotentials are shown by dashed lines. Shocks are marked by
roman digits $I$ and $III$. The accretor is marked by the filled
circle. Vector in the upper right corner corresponds to the
value of velocity of $3A\Omega$.}
\end{figure}

\begin{figure}[t]
\centerline{\hbox{\psfig{figure=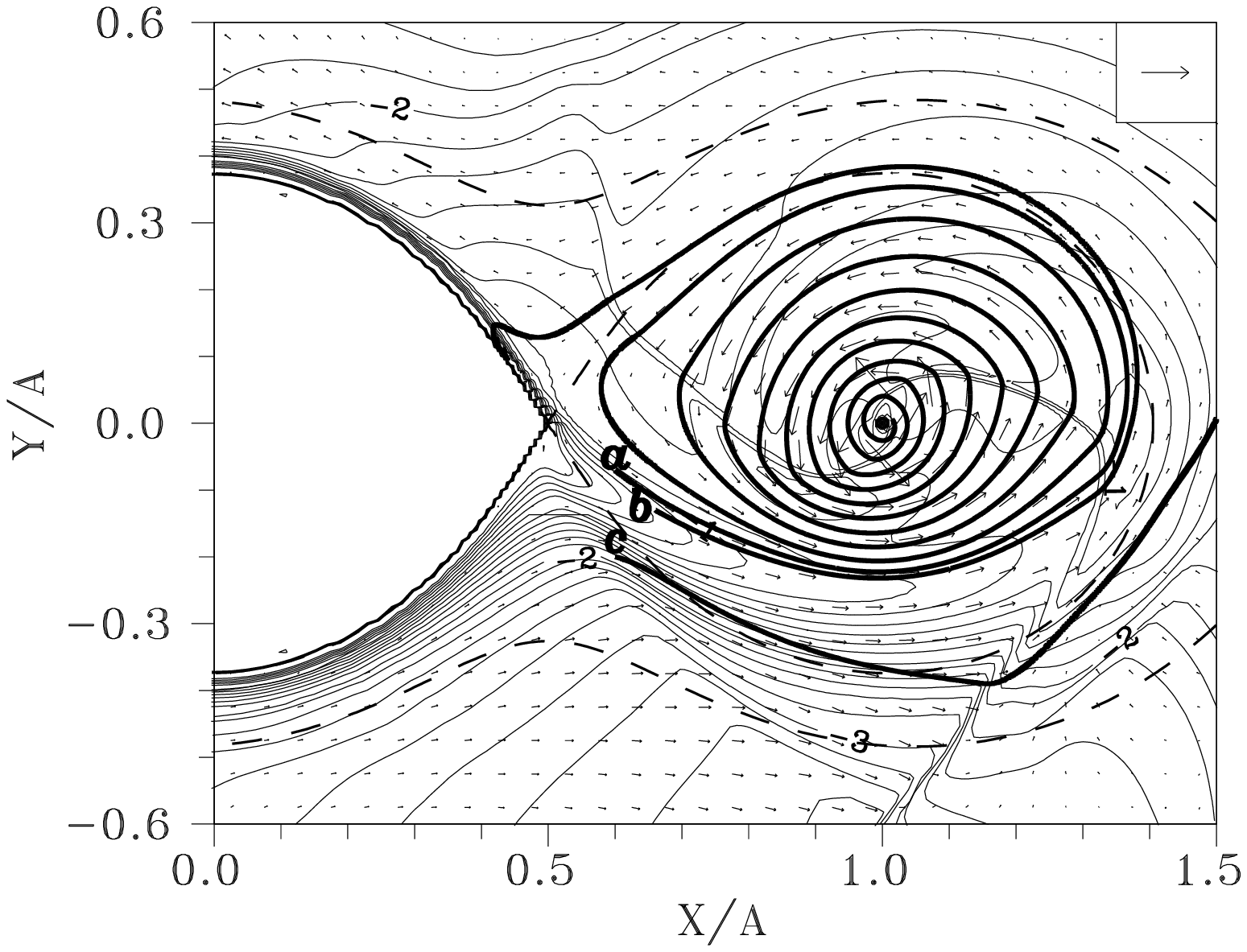,width=6in}}}
\caption{
Density isolines and velocity vectors in
the equatorial plane of the system for the 2D run (case
$\gamma=1.2$) between $10^{-5}..5$ with constant increment of
$\lg \rho$.  Digits on isolines denote values of $\lg \rho$.
Different flowlines are marked by letters "$a$", "$b$", "$c$".
Vector in the upper right corner corresponds to the value of
velocity of $3A\Omega$.}
\end{figure}

\begin{figure}[t]
\centerline{\hbox{\psfig{figure=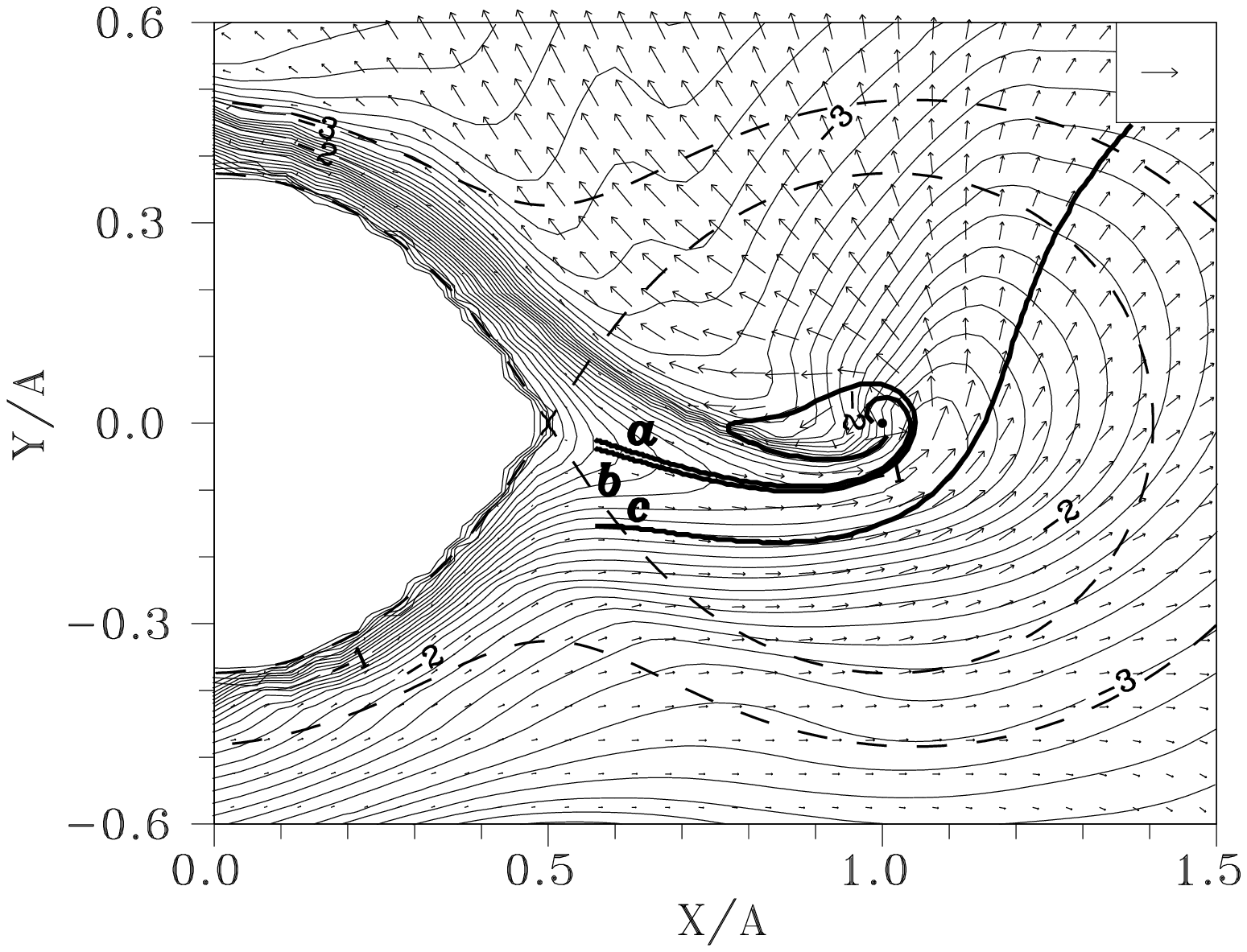,width=6in}}}
\caption{
Density isolines and velocity vectors in
the equatorial plane of the system for the 3D run (case
$\gamma=1.2$) between $10^{-5}..1$ with constant increment of
$\lg \rho$. Digits on isolines denote values of $\lg \rho$.
Different flowlines are marked by letters "$a$", "$b$", "$c$".
Vector in the upper right corner corresponds to the value of
velocity of $3A\Omega$.}
\end{figure}

\begin{figure}[t]
\centerline{\hbox{\psfig{figure=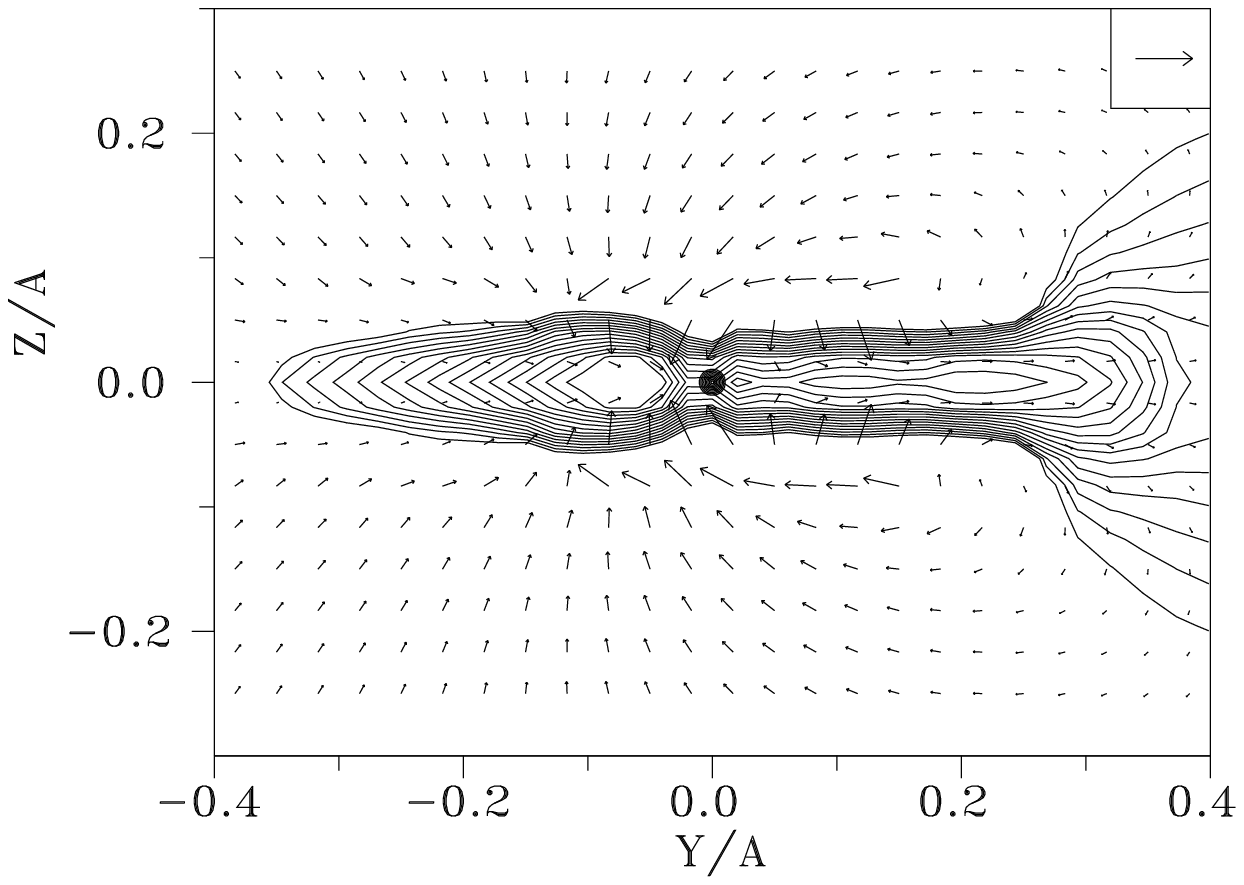,width=6.5in}}}
\caption{
Density isolines and velocity vectors in
$YZ$ plane passing through accretor (case $\gamma=1.01$).
Logarithmic contours are plotted, starting from density of
$0.001..0.1$ (maximum is reached in equatorial plane $z=0$). The
accretor is marked by the filled circle.  Vector in the upper
right corner corresponds to the value of velocity of
$6A\Omega$.}
\end{figure}

\begin{figure}[t]
\centerline{\hbox{\psfig{figure=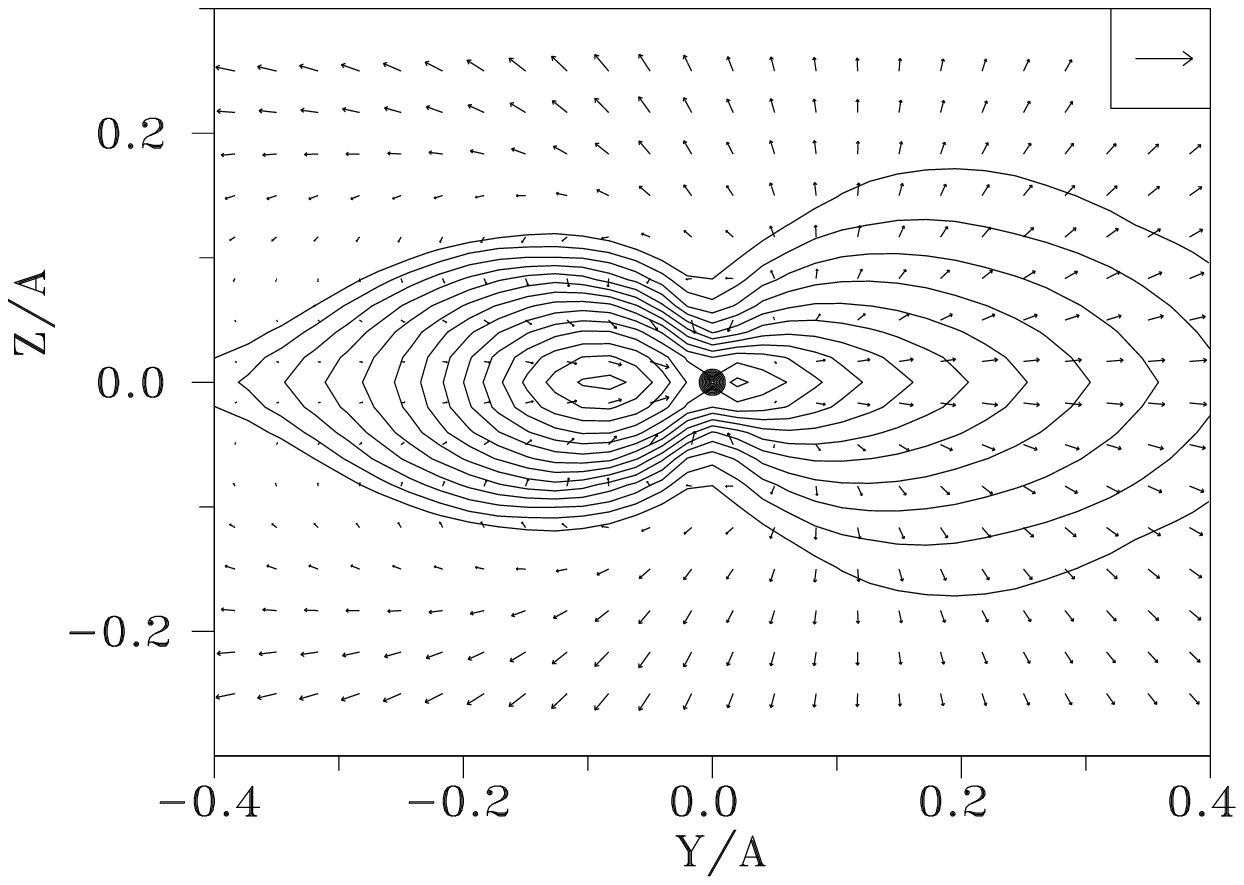,width=6.5in}}}
\caption{
Density isolines and velocity vectors in
$YZ$ plane passing through accretor (case $\gamma=1.2$).
Logarithmic contours are plotted between $0.001..0.1$ (maximum
is reached in equatorial plane $z=0$).  The accretor is marked
by the filled circle.  Vector in the upper right corner
corresponds to the value of velocity of $6A\Omega$.}
\end{figure}

From the above comparison of 2D runs conducted at different
$\gamma$ it is seen that for 2D case the solution does not
change drastically. All characteristic features of the flow
structure (accretion disk, gas fluxes, shocks $I$--$IV$) are the
same, and there are only quantitative changes. It should be also
noted that the described 2D flow structures are close to that
obtained in 2D simulations of Sawada et al. [\ref{SAW86},
\ref{SAW87}].  The observed quantitative changes are not very
significant and can be explained by changing of adopted value of
$\gamma$.

The dependence of the 3D solution from the chosen value of
$\gamma$ is more significant than in 2D case. Comparison of 3D
flow structures obtained at $\gamma$=1.01 (Figs~3, 5) and at
$\gamma$=1.2 (Figs~7, 9) shows the drastic changes in the flow
pattern. The details of these changes will be described in the
next section. Here we simply indicate that for more high value
of $\gamma$ the 3D solution (Figs~7, 9) is different from the 2D
one (Figs~6, 8). In particular, in 3D case the matter of the
stream does not form the accretion disk. Moreover, due to
redistributing of the gas fluxes in the system the only one
spiral-shaped shock $I$ is formed in the system, while the shock
$II$ is disappeared.

Resuming the above points, we may conclude that for
near-isothermal case the 2D and 3D solutions are qualitative
similar. For more high $\gamma$ obtained in 2D and 3D models
solutions are different. This difference is caused by changing
of the 3D flow patterns obtained at different $\gamma$, while
the 2D solutions are comparable and keep all characteristic
features for both considered cases ($\gamma=1.01$ and
$\gamma=1.2$). This conclusion is in a good agreement with study
carried out by Fridman \& Khoruzhii [\ref{FRID94}] where they
have analytically shown that for non-self-gravitational
accretion disk the 2D approach gives an adequate solution only
for isothermal case, while for all other cases the only 3D
models are valid.

\section{3D FLOW STRUCTURE AT DIFFERENT $\gamma$}

As it follows from the comparison of the 3D results obtained for
$\gamma=1.01$ (Figs~3, 5) and $\gamma=1.2$ (Figs~7, 9) the flow
structures in these cases are strongly different. The main
difference is the absence of the accretion disk for the case of
higher $\gamma$.

For more high value of $\gamma$ (Figs~7, 9) the stream goes very
close to accretor. This leads to the direct interaction of the
stream with accretor (flowline "$a$") without forming of the
accretion disk. As a consequence, in steady-state regime for run
with $\gamma=1.2$ the fraction of accreted matter is
approximately equal to 39\% of the total amount of matter
injected into the system by the donor star, while for
near-isothermal case this value is equal to 68\%.
Redistribution of the gas fluxes in the system results in
changing of the positions and intensities of shocks $I$ and
$III$, and disappearing of shock $II$.

As it is seen from the analysis of results the 3D solutions
obtained at different $\gamma$ are qualitative different, while
2D solutions have the same features of the gas flow structure.
Probably this is connected with the changing of the 3D flow
patterns in the third ($Z$) direction. In Figures~10 and 11
density isolines from $\rho \sim 0.1$ up to the value $0.001$
and velocity vectors are shown for cases $\gamma=1.01$ and
$\gamma=1.2$, respectively. These results are presented in the
$YZ$ plane, passing through the accretor and which is
perpendicular to the  line connecting the centers of stars. From
these results it is seen that for near-isothermal case (Fig.~10)
the well defined accretion disk is formed. It is also seen that
in this case the matter moves from the higher altitudes to the
accretor. For case of $\gamma=1.2$ (Fig.~11) the disk is not
formed and the direction of the gas motion is different - the
matter moves to the high altitudes. With the increasing of
specific heats ratio $\gamma$ the compressibility of the gas is
decreased. On the other hand the higher value of $\gamma$ mimics
the case with less cooling efficiency. It means that pressure
and temperature for the $\gamma=1.2$ case are higher than in the
near-isothermal case. In turn, these facts lead to the solution
without accretion disk and gas outflow from the equatorial
plane.

Obtained results are in a good agreement with the 3D simulations
made by Molteni et al. [\ref{MOL91}] and by Lanzafame et al.
[\ref{LANZ92}], where it was found that the accretion disk does
not exist for $\gamma \geq 1.1$. The comparison of the results
of different authors shows that the critical value of $\gamma$
at which the disk does not form depends on the used numerical
scheme and model parameters.

\section{WHAT ARE THE REASONS OF THE\\
SPIRAL-SHAPED SHOCKS FORMATION?}

The results presented above show the formation of the
spiral-shaped shocks $I$ and $II$ in the considered binary. The
obtained flow structures are close to ones described in works
[\ref{SAW86}, \ref{SAW87}] for 2D models and in works
[\ref{SAW92}, \ref{YUK97}] for 3D model.  In a set of previous
works (see, e.g.,
%[\ref{SAW86}, \ref{SAW87}, \ref{SPRU87},
%\ref{ROZ89}, \ref{SAW92}, \ref{YUK97}])
[\ref{SAW86}--\ref{ROZ89},\ref{SAW92},\ref{YUK97}])
these shocks were explained as an intrinsic spirals caused by
tidal disturbances.

The numerical simulations conducted in this work result in
suspicion that the obtained spiral-shaped shocks are due to
colliding of the gas fluxes in the system. The more fine grid
used in our models permits us to consider the details of the
flow structure and to emphasize the features proving the "flow
crossing" nature of these shocks. The analysis of results shows,
that the spiral-shaped shock $I$ is caused by the striking of
the gas of circumbinary envelope revolved around accretor with
the edge of the stream. The second spiral-shaped shock $II$ (in
numerical models where it is present) is also caused by
colliding of the stream with the gas of circumbinary envelope.
The analysis of the gas flow motion presented in Figures~2--9
argues this point.

An additional argument on the "flow-crossing" nature of the
shocks $I$ and $II$ can be obtained from the consideration of
the shocks positions and their lengths. As it is seen from
Figures~2--9 the spiral-shaped shocks are rather long and
propagate far from the accretion disk. The ending points of the
shocks are located in regions where the tidal disturbances are
very small and can not generate the shocks. It is also seen from
the Figures~2--9 the spiral-shaped shocks $I$, $II$ (as well as
non-spiral shocks $III$, $IV$) have the double-shock structures.
This is typical for shocks arising between colliding flows and
can be considered as an decisive prove of the "flow-crossing"
nature of these shocks.

It should be noted that in our simulations we have not found the
formation of the tidally induced intrinsic spiral shocks in the
accretion disk. Probably this is related to the numerical
restrictions (number of grid points, scheme viscosity, etc.) of
the model we used, and it is possible to get a flow structure
containing both "flow-crossing" and intrinsic spiral shocks,
although there are some theoretical indications that at small
$\gamma$ ($\gamma < 1.16$) the formation of spiral shocks is not
possible [\ref{CHAK92}]. From the other hand we can not assert
that these shocks will not arise in simulations of binaries with
different binary parameters.

\section{CONCLUSIONS}

The presented results of systematic comparison of 2D and 3D
numerical models show that only for case of $\gamma \sim 1$ the
obtained solutions are qualitatively similar. For higher value
of $\gamma$ there are strong differences between flow structures
obtained in 2D and 3D models. This means that the 2D models give
a rather adequate results only for near-isothermal case. These
conclusions are in a good agreement with analytical work
[\ref{FRID94}].

The revealed differences between 2D and 3D models are due to the
drastic changing of the 3D flow structure in dependence of
adopted value of $\gamma$, while the 2D models give a
qualitatively similar solutions at different $\gamma$. The main
differences between 3D runs conducted for $\gamma=1.01$ and
$\gamma=1.2$ are the absence of the accretion disk for higher
value of $\gamma$ and the redistribution of the gas flows in the
system.

The spiral-shaped shock waves obtained in presented models are
similar to shocks observed in other numerical simulations (see,
e.g., [\ref{SAW86}, \ref{SAW87}, \ref{SAW92}, \ref{YUK97}]).
In contradiction to previous authors, who  believe that these
shocks are intrinsic spirals caused by tidal disturbances, we
suggest the "flow-crossing" origin of these shocks. The analysis
of the presented results proves this conclusion and shows that
the spiral-shaped shocks are due to collisions of the gaseous
flows in the system.

\section*{ACKNOWLEDGMENTS}

This work was supported by the Russian Foundation for Basic
Research (grant 99-02-17619) and by grant of President of Russia
(99-15-96022).

\section*{REFERENCES}

\small

\begin{enumerate}

\item\label{PRE60}
Prendergast, K.H., {\it Astrophys. J.}, 1960, vol.132, p.162.

\item\label{SAW86}
Sawada, K., Matsuda, T., and Hachisu, I., {\it Mon. Not. R.
Astron. Soc.}, 1986, vol.219, p.75.

\item\label{SAW87}
Sawada, K., Matsuda, T., Inoue, M., and Hachisu, I., {\it Mon.
Not. R. Astron. Soc.}, 1987, vol.224, p.307.

\item\label{SPRU87}
Spruit, H.C., {\it Astron. Astrophys.}, 1987, vol.184, p.173.

\item\label{ROZ89}
R\'o\.zyczka, M., and Spruit, H.C., in {\it Theory of accretion
disks}, ed. F.Meyer et al., Dordrecht: Kluwer, 1989, p.341.

\item\label{TAAM91}
Taam, R.E., Fu, A., and Fryxell, B.A., {\it Astrophys. J.},
1991, vol.371, p.696.

\item\label{BLON95}
Blondin, J.M., Richards, M.T., and Malinowski, M.L.,
{\it Astrophys. J.}, 1995, vol.445, p.939.

\item\label{MUR96}
Murray, J.R., {\it Mon. Not. R. Astron. Soc.}, 1996, vol.279,
p.402.

\item\label{NAG91}
Nagasawa, M., Matsuda, T., and Kuwahara, K., {\it Numer.
Astrophys. in Japan}, 1991, vol.2, p.27.

\item\label{HIR91}
Hirose, M., Osaki, Y., and Minishige, S., {\it Publ. Astron.
Soc. Japan}, 1991, vol.43, p.809.

\item\label{MOL91}
Molteni, D., Belvedere, G., and Lanzafame, G., {\it Mon. Not.
R. Astron. Soc.}, 1991, vol.249, p.748.

\item\label{SAW92}
Sawada, K., and Matsuda, T., {\it Mon. Not. R.
Astron. Soc.}, 1992, vol.255, p.17P.

\item\label{LANZ92}
Lanzafame, G., Belvedere, G., and Molteni, D., {\it Mon. Not.
R. Astron. Soc.}, 1992, vol.258, p.152.

\item\label{BELV93}
Belvedere, G., Lanzafame, G., and Molteni, D., {\it Astron.
Astrophys.}, 1993, vol.280, p.525.

\item\label{MEG93}
Meglicki, Z., Wickramasinghe, D., and Bicknell, G.V., {\it Mon.
Not. R. Astron. Soc.}, 1993, vol.264, p.691.

\item\label{LANZ94}
Lanzafame, G., Belvedere, G., and Molteni, D., {\it Mon. Not.
R. Astron. Soc.}, 1994, vol.267, p.312.

\item\label{BIS97A}
Bisikalo, D.V., Boyarchuk, A.A., Kuznetsov, O.A., and
Chechetkin, V.M., {\it Astron. Zh.}, vol.74, p.880
({\it Astron. Reports}, 1997, vol.41, p.786,
preprint astro-ph/9802004).

\item\label{BIS97B}
Bisikalo, D.V., Boyarchuk, A.A., Kuznetsov, O.A., and
Chechetkin, V.M., {\it Astron. Zh.}, vol.74, p.889
({\it Astron. Reports}, 1997, vol.41, p.794,
preprint astro-ph/9802039)

\item\label{YUK97}
Yukawa, H., Boffin, H.M.J., and Matsuda, T., {\it Mon.
Not. R. Astron. Soc.}, 1997, vol.292, p.321.

\item\label{BIS98}
Bisikalo, D.V., Boyarchuk, A.A., Chechetkin, V.M., Kuznetsov,
O.A., and Molteni, D., {\it Mon. Not. R. Astron. Soc.}, 1998,
vol.300, p.39.
%preprint astro-ph/9805621

\item\label{FRID94}
Fridman, A.M., and Khoruzhii, O.V., in Gor'kavyi,
N.N., Fridman, A.M., {\it Physics of planetary rings},
Moscow: Nauka, 1994, p. 282 (in Russian).

\item\label{ROE86}
Roe, P.L., {\it Ann. Rev. Fluid Mech.}, 1986, vol.18, p.337.

\item\label{CHAK85}
Chakravarthi, S., and Osher, S., {\it AIAA Pap.}, 1985, N
85--0363.

\item\label{EIN88}
Einfeldt, B., {\it SIAM J. Numer. Anal.}, 1988, vol.25, p.294.

\item\label{LANDAW}
Landau, L.D., and Lifshitz, E.M., {\it Fluid mechanics},
Elmsford: Pergamon Press, 1959.

\item\label{CHAK92}
Chakrabarty, S.K., {\it Mon. Not. R. Astron. Soc.}, 1992,
vol.259, p.410.

\end{enumerate}

\end{document}